\documentclass[ps,prd,preprintnumbers,superscriptaddress,nofootinbib,floatfix,twocolumn,notitlepage]{revtex4-2}
\pdfoutput=1 

\usepackage{dcolumn}

\usepackage{bm,amssymb,slashed,graphicx,multirow,soul,mathtools,xspace,array,tikz,amsmath,siunitx}
\usepackage[compat=1.1.0]{tikz-feynman}   
\usepackage{float}                   
\allowdisplaybreaks
\usepackage{ bbold }
\usepackage{caption,subcaption}
\usepackage{braket}
 \usepackage{hyperref}

\definecolor{nicered}{rgb}{0.7,0.1,0.1}
\definecolor{nicegreen}{rgb}{0.1,0.5,0.1}
\definecolor{violet}{rgb}{0.7,0.3,0.3}
\hypersetup{colorlinks,citecolor= nicegreen,linkcolor= nicered}

\newcommand{\lp}{\left(}
\newcommand{\rp}{\right)}

\newcommand{\beq}{\begin{equation} }
\newcommand{\eeq}{\end{equation}} 
\newcommand{\bi}{\begin{itemize} }
\newcommand{\ei}{\end{itemize} }

\definecolor{Red}{rgb}{1.,0.,0.}
\definecolor{Grn}{rgb}{0.,0.75,0.}
\definecolor{Blu}{rgb}{0.,0.,1.}
\definecolor{Pink}{rgb}{1,0.08,0.58}

\usepackage[T1]{fontenc} 

\usepackage{amsmath,amssymb,epsfig,ulem,color,slashed}
\allowdisplaybreaks  

\newcommand{\Xm}{X_{\rm max}}

\begin{document} 


\title{\boldmath Improving Composition of Ultra High Energy Cosmic Rays with Ground Detector Data }


\author{Bla\v{z} Bortolato}
 \email{blaz.bortolato@ijs.si}
\author{Jernej F. Kamenik}
 \email{jernej.kamenik@cern.ch}
\affiliation{
 Jo\v{z}ef Stefan Institute, Jamova 39, 1000 Ljubljana, Slovenia \\
 Faculty of Mathematics and Physics, University of Ljubljana, Jadranska 19, 1000 Ljubljana, Slovenia
}%

\author{Michele Tammaro}
 \email{michele.tammaro@ijs.si}
\affiliation{
 Jo\v{z}ef Stefan Institute, Jamova 39, 1000 Ljubljana, Slovenia
}%

\date{\today}

\begin{abstract}
We show that the maximum shower depth ($\Xm$) distributions of Ultra-High Energy Cosmic Rays (UHECRs), as measured by fluorescence telescopes, can be augmented by building a mapping to observables collected by surface detectors. 
The resulting statistical improvement of such augmented dataset depends in a universal way on the strength of the correlation exhibited by the mapping. 
Building upon the publicly available data on ``golden hybrid'' events from the Pierre Auger Observatory we project possible improvements in the inferred composition of UHECRs for a range of possible mappings with varying correlation strengths. 
\end{abstract}

\maketitle

\section{Introduction}
\label{sec:intro}
Cosmic rays (CRs) with ultra-high energy, $E \geq 10^{18}$ eV, are studied through the observation of the Extensive Air Showers (EAS) that are generated in the Earth atmosphere. The Pierre Auger Observatory~\cite{PierreAuger:2015eyc} employs a ``hybrid'' detection method: Surface Detectors (SD) measure electrons and muons propagating to the surface, while Fluorescent Detectors (FD) capture the isotropic light emitted by excited nytrogen and can track the development of the EAS. In particular, the former can measure the atmospheric depth at which the number of charged particles is at its maximum, $\Xm$, which is in turn a key ingredient to infer the mass composition of CRs.

While the hybrid measurement technique provides detailed insights for the EAS study, the runtime of FDs is largely limited by background light noise. This results in $\sim10\%$ of the total number of showers observed to be reconstructed as ``hybrid'', i.e. with both FD and SD data, while the remaining part of the dataset is measured with the SD only. For example, in the Pierre Auger 2021 Open Data~\cite{the_pierre_auger_collaboration_2021_4487613}, there are 22731 SD measurements of EAS, which
we refer to as Non-Hybrid (NH) showers, and 3156 “brass hybrid” (BH) events, that is showers that have been recorded simultaneously by the SD and the FD. Of these BH, 1602 are called “golden hybrids” (GH), with independent SD and FD reconstructions. The low statistics of BH and especially GH events compared to NH is currently a dominant limiting factor in the determination of the CR mass composition.

A possible solution is to infer $\Xm$ from ground data, exploiting GH events to build a map between SD and FD observables. Unfortunately, linear and simple non-linear fits based solely on measured data, such as the $\Delta$ variable proposed in Ref.~\cite{PierreAuger:2017tlx}, exhibit only relatively low levels of predictive power. Recently, an alternative has been proposed in terms of a Deep Neural Network trained on simulated ground data for 4 different primaries to directly predict $\Xm$~\cite{PierreAuger:2021fkf}. 
Unfortunately, simulations currently cannot faithfully reproduce the observed ground data of Pierre Auger events~\cite{Cazon:2020zhx}. This exposes any approach directly relying on such simulations to potentially large systematic errors. In addition, the chosen composition of simulated events used to train the network can act as a confounding variable on the inferred $\Xm$. This can in turn  lead to biased estimates of CR composition. 
 
In this work we address these drawbacks and propose a general method to exploit any observed correlations between SD and FD data, to effectively increase the statistics of $\Xm$. After briefly reviewing possible maps in Section~\ref{sec:Correlation}, in Section~\ref{sec:Inference} we show a rigorous way to combine the GH data with the inferred one, which leads to the enhancement of the dataset by an effective number of GH showers, $N_{{\rm eff}} = f \times N_{{\rm GH}}$. The latter number monotonically increases with the level of correlation given by the map. Finally, in Section~\ref{sec:Results} we employ this method to improve the results on the full mass composition obtained in Ref.~\cite{Bortolato:2022ocs}.

\section{Correlating ground detector data with $\Xm$}
\label{sec:Correlation}

For our purpose the $i$-th GH event in the dataset can be characterized by one FD measurement, $x^{(i)} \equiv \Xm^{(i)}$, and a set of $m$ SD observables, ${\cal Q}^{(i)} = \{Q_1^{(i)},\dots,Q_m^{(i)} \}$. We can then consider functions of SD observables $y(\{Q\})$ for which the GH set exhibits some correlation between $x$ and $y$, as parametrized by
\beq
\rho(x,y) = \frac{\text{Cov}(x,y)}{\sigma(x)\sigma(y)}\,,
\eeq
where $\sigma(x)$, $\sigma(y)$ and ${\rm Cov}(x,y)$ are the standard deviations of $x$ and $y$ and their covariance, respectively. 

Possibly the simplest example of $y$ is the weighted average of $Q_j$
\beq
\label{eq:y_variable}
y^{(i)} = \sum^m_{j = 1}  a_j Q_j^{(i)}\,,
\eeq
where the weights $a_j$ are fixed by maximizing $\rho(x,y)^2$. For simplicity, we assume that the PDF of each variable is a normal distribution described by its measurement and the respective uncertainty. Namely, we have
\beq\label{eq:PDFvariable}
    P^{(i)}(x) \sim {\cal N}\lp x ~|~ x^{(i)},\delta x^{(i)} \rp \,, 
\eeq
and similarly for $P^{(i)}(y)$.
Note that the fit has to be performed separately for each subset of GH events (i.e. in a particular CR energy bin or angular window) for which one wants to infer composition. This ensures an unbiased estimator, assuming the GH and NH events in the same subset are of the same composition and follow the same distribution of ${\cal Q}^{}$.

As a proof of concept, we use the set of observables\footnote{We have repeated this exercise with multiple combinations of available observables. The chosen set represents the one with the largest correlation achievable without overfitting of the data.}
\beq\label{eq:SDobservables:Proof}
{\cal Q}^{(i)} = \{\theta,~\langle T_s \rangle, ~\sigma(T_s), ~\max(T_s) \}^{(i)}\,,
\eeq
with their respective uncertainties. Here, $\theta$ is the azimuthal angle of the shower core, while $T_s = (t_{{\rm end}} - t_{{\rm start}} )_s$ is the duration of the signal in the $s$-th Cherenkov station hit by the shower, defined as the difference between the final and initial time bin. 
We take the mean, standard deviation and maximum values over the active stations for each event. 
For each shower, we generate multiple samples of the observables from their PDF, eq.~\eqref{eq:PDFvariable}. In such a way, we can also compute the uncertainty associated with $y^{(i)}$. Uncertainties on the bins $t_{{\rm start}}$ and $t_{{\rm end}}$ are not provided in the Open Data, hence we assume the uncertainties on both quantities to be $\pm$ one bin.
These samples are then utilized to maximize $\rho^2$ through the ``Nelder-Mead'' method (any similar optimization algorithm would work). Finally, we employ $K$-folding to check for overfitting.
Following this procedure, we are able to achieve $\rho = (30\pm 4)\%$ correlation between $x$ and $y$ for GH in the energy bin $[2,~5]$ EeV, with 455 GH showers. 
We chose this bin as the only one where GH and NH data overlap in the Pierre Auger Open Data set.

In practice, more advanced approaches, such as non-linear fits or neural networks, may be employed to obtain stronger correlations between $y$ and $x$. As two explicit examples of such more complicated observables, we mention the $\Delta$ variable~\cite{PierreAuger:2017tlx} and the Deep Neural Network trained on simulated ground data~\cite{PierreAuger:2021fkf}. In the same energy bin, the former yields  $\rho = 0.26 \pm 0.02$~\footnote{Uncertainties due to limited statistics and above stated systematics in both cases have been estimated using bootstrapping~\cite{EfroTibs93}.}, while the latter can reach $\rho=63\%$. We stress that for our purpose the strength of an observable is measured solely in terms its correlation with $x$ {\it as established on a particular measured GH dataset} $\{(x,y)\}$. In this way we avoid potential systematic biases associated with direct inference of individual $x^{(i)}$ from non-equivalent datasets or simulations. 

\section{Inference of $\Xm$ distribution}
\label{sec:Inference}
Following the previous Section, we can represent the GH dataset as a set of $N$ pairs $(x_j, y_j)$, with $j=1,\dots,N$, and we assume that there is some level of correlation between the two observables. The NH dataset is then represented by the $M$ pairs ${(\hat x_k, \hat y_k)}$, with $k=1,\dots,M$, where $M \gg N$. In the latter set the $\hat x_k$ entries, which are missing in the original data, can be inferred by exploiting the map built through the $\{(x,y)\}$ set correlation, that is $\hat x = \hat x(\hat y)$.

Let $P(x,y)$ be the joint probability distribution of the $\{(x,y)\}$ set. For simplicity, we start by assuming that such distribution follows a Bivariate Normal (BN) model, while we comment later for the general case. Each BN distribution is defined by the means $\mu_{ x}$, $\mu_{ y}$, the standard deviations $\sigma_{ x}$, $\sigma_{ y}$ and the correlation $\rho$. The conditional distribution $P( x |  y)$ is then simply given by the normal distribution with mean $\mu_{ x} + \rho \frac{\sigma_{ x}}{\sigma_{ y}} ( y - \mu_{ y})$ and standard deviation $\sigma_{ x} \sqrt{1 + \rho^2} $. Therefore one can sample  $\hat x \sim P( x |  y)$ following the formula
\beq
\label{eq:xhat}
    \hat x(\hat y) = \mu_{ x} + \rho \frac{\sigma_{ x}}{\sigma_{ y}} (\hat y^\prime - \mu_{ y}) +  
    \sigma_{ x} \sqrt{1 + \rho^2} ~ \epsilon\,,
\eeq
where $\hat y^\prime = \hat y + \delta \hat y~\epsilon$ accounts for systematic uncertainty of $\hat y$, and $\epsilon$ is a random number drawn from a normal distribution, $\epsilon \sim \mathcal{N}(0,1)$. In this way we obtain a set of inferred values, $\{\hat x(\hat y_1), \dots, \hat x(\hat y_M) \}$.

Due to the finite number ($N$ and $M$) of events, we need to account for statistical uncertainties in the inference via bootstrapping~\cite{EfroTibs93}. The latter method consists of inferring values $\hat x$ for each $\hat y$ multiple times, each time starting with a different set of pairs $\{(\tilde{x}, \tilde{y})\}$, called the bootstrapped sample. The procedure is repeated $B$ times, where each time the bootstrapped sample is given by sampling $N$ pairs $(\tilde x_j, \tilde y_j)$ from the first set $\{(x,y)\}$ with allowed repetitions. 

Namely, at bootstrap step $\ell$, where $\ell=1,\dots,B$, we have
\beq
\begin{split}
\tilde{x}_{j,\ell} &= \lp x_j + \delta x_j \epsilon_{j,\ell} \rp {\cal O}_{j,\ell}\,,\\ 
\tilde{y}_{j,\ell} &= \lp y_j + \delta y_j \epsilon_{j,\ell} \rp {\cal O}_{j,\ell}\,.
\end{split}
\eeq
Here $\delta x_j$ and $\delta y_j$ represent the systematic uncertainties on $x_j$ and $y_j$, while $\epsilon_{j,\ell}\sim \mathcal{N}(0,1)$. Finally, the $N$ dimensional vector ${\cal O}_{\ell}$ is drawn from the Multinomial distribution with $N$ events and $N$ classes which are all equally probable, $\text{Mult}(N, p = (1/N,...,1/N))$. The values in this array, ${\cal O}_{j,\ell}$, count how many times each pair $j$ is chosen in the $\ell$-th bootstrapped sample. It thus satisfies the relation  $\sum_j {\cal O}_{\ell, j} = N$. As a result of this procedure, we obtain at each step $\ell$ the set of $N$ couples
\beq
\mathcal{X}_{\ell} = \{(\tilde x, \tilde y) \}_\ell\,.
\eeq
We can now estimate the $\ell$-th joint distribution $P_\ell(\tilde x, \tilde y)$ from the bootstrapped samples.

Similarly, we define
\beq
\tilde{\hat y}_{k,\ell} = (\hat y_k + \delta \hat y_k \epsilon_{k,\ell}) \tilde{\mathcal{O}}_{k,\ell}\,,
\eeq
where now the index $k$ runs over the $\hat y$ elements, $k=1,\dots,M$, and $\tilde{\mathcal{O}}_{k,\ell}\sim\text{Mult}(M, p = (1/M,...,1/M))$. 
The $\tilde{\hat y}_{k,\ell}$ account for both statistical and systematic uncertainties. 

\begin{figure}[h!]
    \centering
    \includegraphics[scale = 0.85]{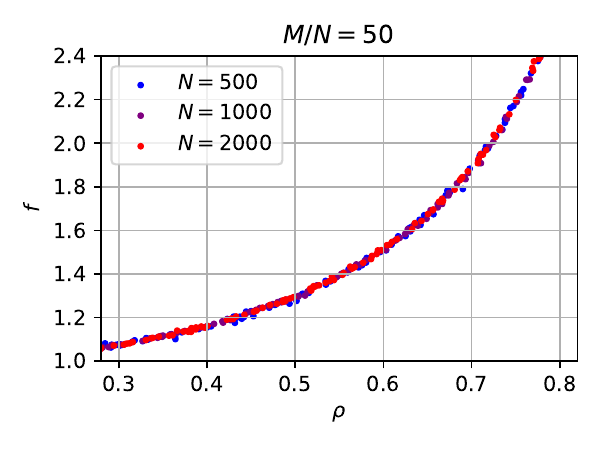}
    \caption{Effective increase $f$ as a function of correlation $\rho$, for three values of GH dataset size $N$, when fixing the ratio $M/N = 50$.}
    \label{fig:plot_r50}
\end{figure}

\begin{figure}[h!]
    \centering
    \includegraphics[scale = 0.85]{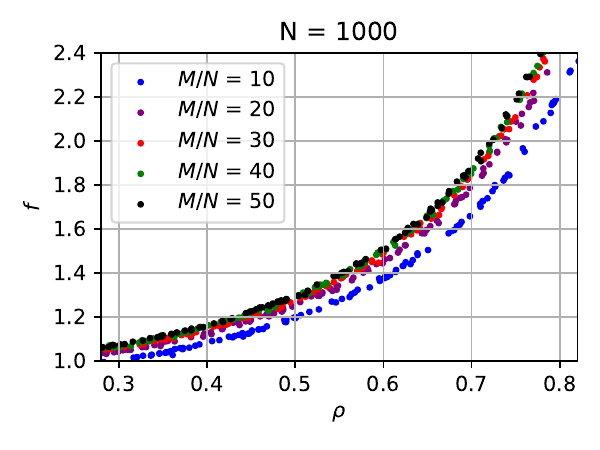}
    \caption{Same as Fig.~\ref{fig:plot_r50}, with $N=1000$ and different values of the ratio $M/N$.}
    \label{fig:plot_diff_ratios}
\end{figure}

For each bootstrapped sample we can now obtain a set of $M$ pairs 
\beq
{\cal X}_\ell^{{\rm inf}} = \{ (\hat{x}(\tilde{\hat{y}}), ~ \tilde{\hat y}) \}_\ell\,,
\eeq
where the first component of the pair is inferred by plugging $\tilde{\hat y}_{k,\ell}$ in Eq.~\eqref{eq:xhat}. Finally, we can combine the two sets to obtain
\beq
\mathcal{X}^{\rm comb}_l = {\cal X}_\ell  \cup {\cal X}_\ell^{{\rm inf}} = \{ (X, Y) \}_\ell\,.
\eeq
At this point, one can use the larger combined dataset to perform the desired analysis. As we obtain the $\hat x$ elements from the inference method described above, it is fair to ask what is the gain in terms of statistical power with the combined dataset. In order to estimate it, we can compute the variance of the distribution of the means, ${\rm Var}(\bar x)$. That is, we compute the mean value of $\tilde x$ for each ${\cal X}_\ell$ set, $\bar x_\ell$; the resulting distribution defined by all $\bar x_\ell$ is a Gaussian with variance ${\rm Var}(\bar x)$. 
The ratio between the variances obtained with the initial dataset ${\cal X}_\ell$ and with the combined one, ${\rm Var}(\bar X)$, can provide a measure of the improvement in the statistical significance obtain with this method. Indeed, the quantity
\beq
f  \equiv \frac{\text{Var}(\bar x)}{\text{Var}(\bar X)}\,,
\eeq
can be interpreted as the effective increase of events in the combined dataset, with respect to the original one, $N_{\rm eff} = f\times N$. This interpretation follows from the fact that the variance of the mean distribution is given by the variance divided by the number of events. Namely, if $f = 1$ it means no additional information is contained in the inferred set ${\cal X}^{\rm inf}$. On the contrary, if $f>1$ the effective statistical power of ${\cal X}^{\rm comb}$ amounts to have $f\times N$ total events.
The functional dependence of $f$ on the correlation $\rho$, for different values of $M$ and $N$, is shown in Figs.~\ref{fig:plot_r50} and \ref{fig:plot_diff_ratios}. Points in these figures are obtained numerically, with the procedure described above, using $B = 10^5$ bootstrapped samples. Furthermore, we assumed that systematic uncertainties are small. The latter translates to have $\rho \sigma_x / \sigma_y \delta \hat y \ll \sigma_x \sqrt{1 - \rho^2}$ in Eq.~\eqref{eq:xhat}. In this scenario, the ratio $f$ increases monotonically with $\rho$ and does not depend on the number of events $M$ and $N$, but only on their ratio, see Fig.~\ref{fig:plot_r50}. Larger $M/N$ ratios allow then for larger values of $f$ for the same level of correlation, as shown in Fig.~\ref{fig:plot_diff_ratios}. 

In the limit of $\rho \rightarrow 1$, $f$ tends to the maximal value $f = (M+1)/N$. In practice, $\rho = 1$ cannot be obtained with a finite number of events $N$ due to statistical uncertainties, thus the maximum value of $f$ is smaller.

In the general case, the joint distribution $P(x,y)$ does not follow a BN distribution. One may then try to use Kernel Density Estimation (KDE) to obtain $P(x,y)$; however, if the latter is computed inappropriately, we may introduce a bias in inferring $\hat x$ values (variance - bias trade-off). For this reason we propose to transform bootstrapped samples using a Probability Integral Transformation in 2D \cite{3a86ccbf-0ae5-3066-9917-815106cd0e39}, which has the advantage to preserve the correlation between $x$ and $y$. 

In the Auger Open Data, NH and GH events given in the Open Data in the energy interval $E/\rm{EeV} \in [2.5, \ 10]$ amount to $M\sim 15700$ and $N\sim 300$ respectively, thus with a ratio $M/N\sim50$. In the case of maps between $y$ and $x\equiv\Xm$ constructed with linear or non-linear fits (see previous Section), the correlation is $\rho \simeq 0.3$, which corresponds to $f \approx 1.06$. 
On the other hand, the Deep Neural Network presented in Ref.~\cite{PierreAuger:2021fkf} can reach $\rho = 63\%$, resulting in $f = 1.4 - 1.5$ for $M/N = 10 - 50$.

\section{Projections of UHECR Composition from non-hybrid events}
\label{sec:Results}
Using the $\Xm$ distribution inferred from a combination of GH and NH events, we follow our previous analysis of the mass composition of UHECR~\cite{Bortolato:2022ocs}. That is, we infer the posterior distribution of the composition $w = (w_p,\dots,w_{{\rm Fe}})$ using the moment decomposition of measured and simulated $\Xm$ distributions in a fixed energy bin. However, due to the small number of events in the energy bin $E/{\rm EeV} \in [2.5, 5]$, we use here the larger $E/{\rm EeV} \in [0.65, 5]$ energy bin for projections. We use the same set of shower simulations and the same inference procedure of Ref.~\cite{Bortolato:2022ocs}; the improvements shown here come solely from considering the enhanced dataset, $\mathcal{X}_\ell^{\rm comb} \equiv \{ X_{\ell,k} |~ k = 1,...,N+M\}$.

At each bootstrap step $\ell$, we compute the moments of the $\Xm$ distribution as
\begin{align}
z_{\ell,1} &= \frac{1}{N+M} \sum^{N+M}_{k = 1} X_{\ell,k}\, \\
z_{\ell,n} &= \frac{1}{N+M} \sum^{N+M}_{k = 1} \left( X_{\ell,k} - z_{\ell,1} \right)^n\,
\end{align}
for $n=2,3,4$. We then calculate the mean of the distribution of moments $\mu$ and the covariance matrix $\Sigma$. The effect of the enhanced dataset can be included here by considering the new covariance matrix $\Sigma_{\rm eff} = \Sigma/f$. 

\begin{figure}[h!]
    \centering
    \includegraphics[scale = 0.8]{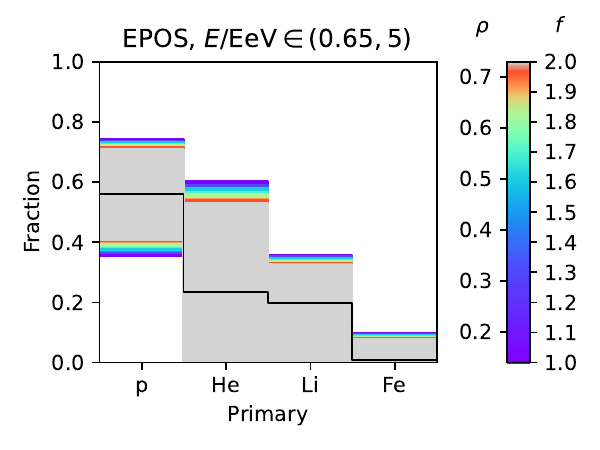}
    \caption{Fraction of primaries p, He, Li, Fe shown as $2\sigma$ confidence intervals for various values of $f$ and corresponding $\rho$ for hadronic model EPOS. Black solid line represent the most probable composition.}
    \label{fig:EPOS_4primaries}
\end{figure}

In Fig.~\ref{fig:EPOS_4primaries} we show the composition obtained using a mixture model of 4 primaries, (p, He, N, Fe), using the EPOS hadronic model. The black line indicates the best fit, while the bars represent the allowed fractions at 95\% CL. Different colors show the effect of including inferred values of $\Xm$, for different values of $f$: the grey band obtained for $f=2$, that is by doubling the size of the GH dataset, down to the purple one, where $f=1$.

\begin{figure}[h!]
    \centering
    \includegraphics[scale = 0.8]{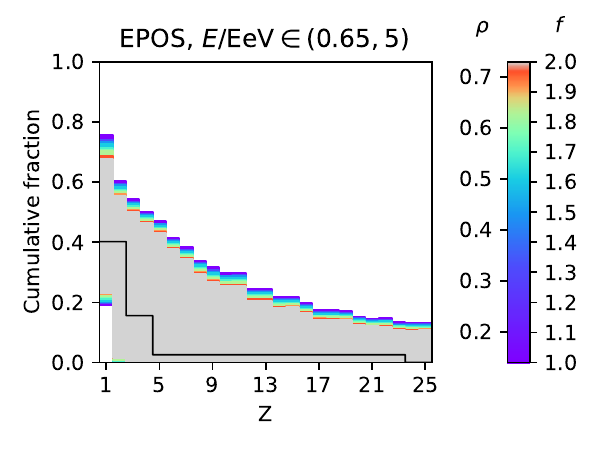}
    \caption{Fraction of primaries shown as $2\sigma$ confidence intervals for various values of $f$ and corresponding $\rho$ for hadronic model EPOS. Black solid line represent the most probable composition.}
    \label{fig:EPOS_26primaries}
\end{figure}

Similarly, in Fig.~\ref{fig:EPOS_26primaries} we show the result for the full cumulative composition, that is for each primary $Z_0$ we show the allowed fraction of heavier primaries, $Z>Z_0$. We observe again an improvement of the grey region with respect to the purple. 

We provide the same results for the hadronic models Sibyll, QGSJet01 and QGSJetII-04 in the Appendix.

\section{Conclusions}
\label{sec:Conclusions}
In this Letter we presented a novel method to improve the statistical power of GH events by including $\Xm$ values of UHECRs inferred from SD observables. The effective number of events obtained is quantified by the variance ratio $f$, which strongly depends on the level of correlation between ground and fluorescent data. While simple linear or non-linear fits can achieve at most $30\%$ correlation levels, resulting in rather modest values of $f$, more advanced approaches can reach above $60\%$ correlation~\cite{PierreAuger:2021fkf}, leading to $f\gtrsim1.5$. 

We then project the CR composition inference with such  enhanced datasets, as a function of the ratio $f$. 
The resulting projected improvements in the composition CL for the Auger Open Data are shown in Figs~\ref{fig:EPOS_4primaries} and~\ref{fig:EPOS_26primaries} and in the Appendix.
Although the latter results seems meager, this study shows that, as a proof-of-concept, these improvements are possible and can be exploited by the experimental collaboration on their full dataset. Furthermore, it underlines the potential of probing deeper the correlation of SD and FD data, where the advancement in artificial intelligence tools can lead to interesting and powerful results. 

Finally, we point out that the procedure to enhance the $\Xm$ dataset by SD observables is very general and based on statistics arguments only. That is, any analysis based on correlated observables can make use of this method to augment a dataset containing measurements of both observables with (preferably a much larger)  dataset of measurements of only a single one.\footnote{For Pierre Auger, this could be advantageous i.e. in the exploration of observables to discriminate between hadronic models, a direction we leave for future work.} 


\section*{Acknowledgments}
We are grateful to Gregor Šega for helping in reviewing the method outlined in the Letter.
The authors acknowledge the financial support from the Slovenian Research Agency (grant No. J1-3013 and research core funding No. P1-0035).

\clearpage

\onecolumngrid
\begin{appendix}
\section{UHECR COMPOSITION FROM NON-HYBRID EVENTS}
\label{app:Additional_results}
\begin{figure}[h!]
    \centering
    \includegraphics[scale = 0.9]{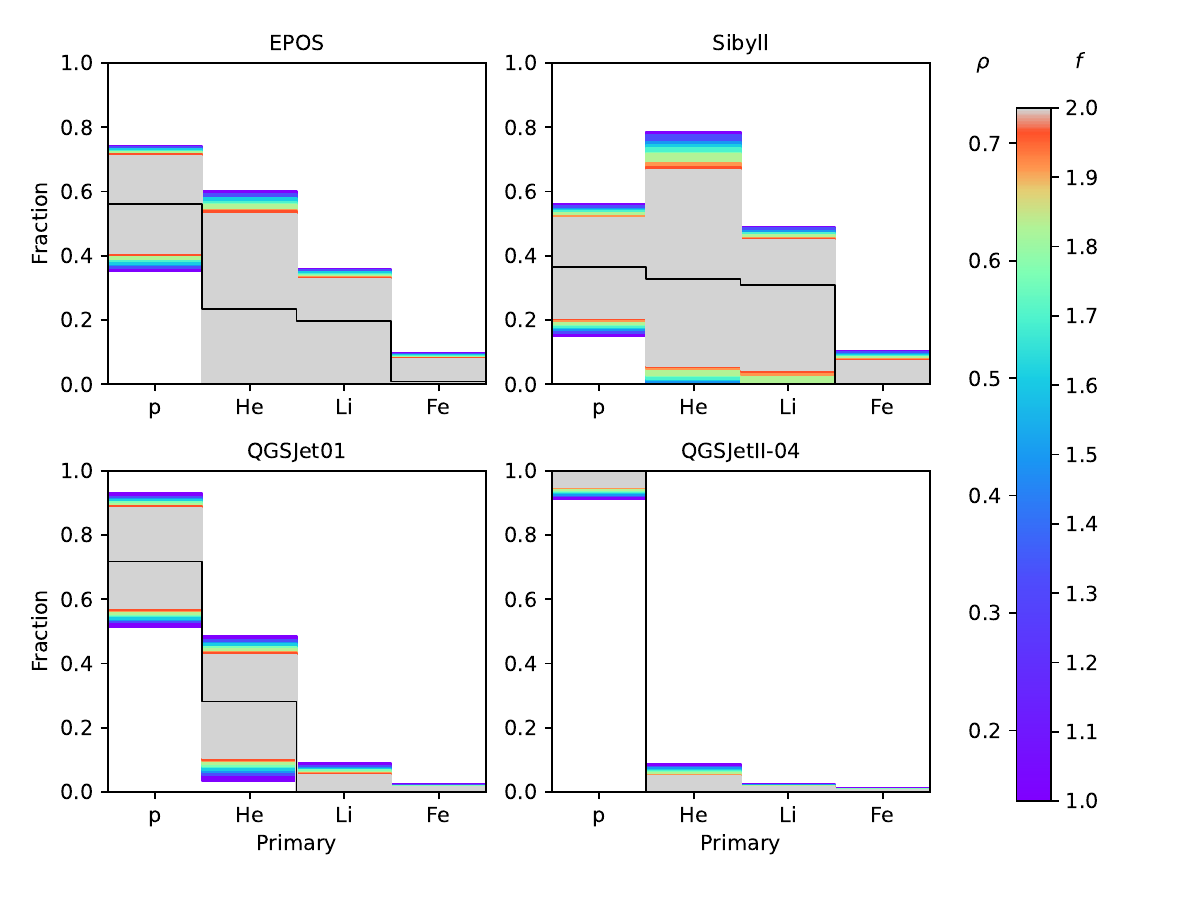}
    \caption{Fraction of primaries p, He, Li, Fe shown as $2\sigma$ confidence intervals for various values of $f$ and corresponding $\rho$ for hadronic models EPOS, Sybill, QGSJetII-04 and QGSJet01 in energy interval $E/{\rm EeV} \in [0.65, 5]$. Black solid line represent the most probable composition.}
    \label{fig:Projection_4primaries}
\end{figure}

\begin{figure}[h!]
    \centering
    \includegraphics[scale = 0.9]{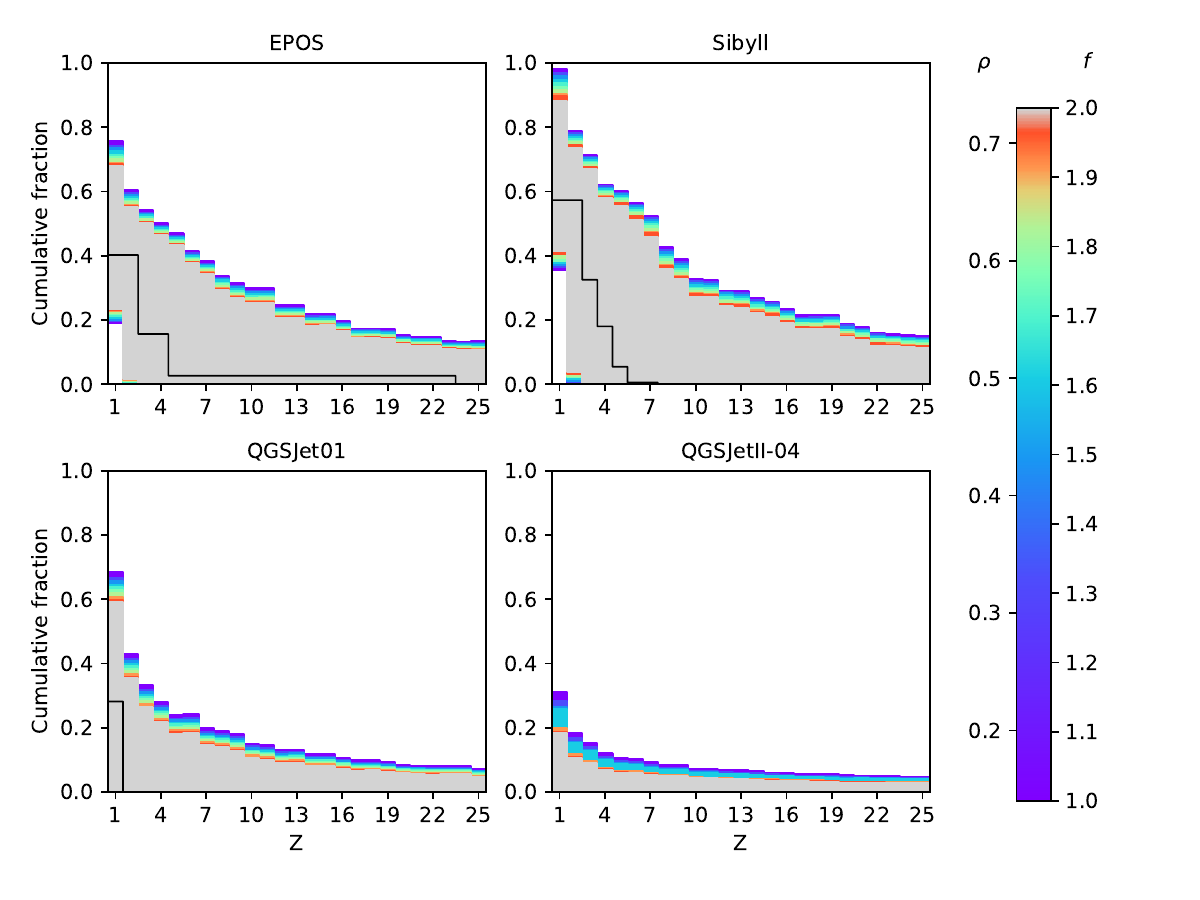}
    \caption{Fraction of primaries shown as $2\sigma$ confidence intervals for various values of $f$ and corresponding $\rho$ for hadronic models EPOS, Sybill, QGSJetII-04 and QGSJet01 in energy interval $E/{\rm EeV} \in [0.65, 5]$ . Black solid line represent the most probable composition.}
    \label{fig:Projection_26primaries}
\end{figure}

Below we present the projections of compositions for all hadronic models considered: EPOS (as in the main text), Sybill, QGSJetII-04 and QGSJet01 in the energy bin $E/{\rm EeV} \in [0.65, 5]$.
In case of 4 primaries projections are shown in Fig.~\ref{fig:Projection_4primaries},
while for 26 primaries are given in Fig.~\ref{fig:Projection_26primaries}.

\end{appendix}
\clearpage

\bibliographystyle{h-physrev}
\bibliography{references}

\end{document}